# Unimodular Supergravity

Hitoshi Nishino[1] and Subhash Rajpoot[2]

*Department of Physics & Astronomy*
*California State University*
*Long Beach, CA 90840*

## Abstract

We present the locally supersymmetric formulation of unimodular gravity theory in $D$ $(1 \leq D \leq 11)$ dimensions, namely supergravity theory with the metric tensor whose determinant is constrained to be unity. In such a formulation, the usual fine-tuning of cosmological constant is no longer needed, but its value is understood as an initial condition. Moreover, the zero-ness of the cosmological constant is concluded as the most probable configuration, based on the effective vacuum functional. We also show that the closure of supersymmetry gauge algebra is consistent with the unimodular condition on the metric.

---

[1]E-Mail: hnishino@csulb.edu

[2]E-Mail: rajpoot@csulb.edu

## 1. Introduction

Ever since Einstein's 'blunder' [1], how to understand the zero or extremely small cosmological constant without fine-tuning has been a long-standing problem both at the classical and quantum levels [2]. Lagrangian formulation of general relativity admits it, no known symmetry forbids it, and up until recently, it was not even required empirically. Recent Type Ia supernova observation [3] provides evidence that the universe is accelerating at a greater rate now than in the past, and implies a non-zero cosmological constant $(\Lambda \neq 0)$. An interesting implication of this is that energy density $\Omega_\Lambda$ associated with non-zero $\Lambda$ is of the same order of magnitude as the matter density of the universe, giving rise to so-called second cosmological problem. This has led to a flurry of activity explaining the two cosmological problems, and involves the anthropic principle [2][4], quintessence [5], new interactions, extra dimensions, phase transitions, and space-time fluctuations. However, more data are required before definite conclusions can be drawn. Here we address only the 'first' cosmological constant problem.

In a certain formulation, the cosmological constant problem can be understood as an 'initial condition' instead of extremely small number adjusted by hand as an artificial 'fine-tuning'. Such a theory is called 'unimodular gravity' theory, in which the determinant of the metric tensor is constrained to be unity, originally developed in [6][7]. Motivated by the development of a possible solution to the cosmological constant based on baby universe with wormholes [8], the authors in [9] computed the effective vacuum functional in unimodular gravity theory as

$$Z = \int d\mu(\Lambda) \, \exp\left(\tfrac{3\pi}{G\Lambda}\right) \quad , \tag{1.1}$$

where $G$ is the Newton's constant, and $d\mu(\Lambda)$ is the path-integral measure for the 'scalar field' $\Lambda(x)$. Even though $\Lambda$ is initially a scalar field, it is constrained to be space-time independent by a lagrange multiplier field. Similarly to wormhole models [8], this $Z$ has a singularity at $G^2\Lambda = 0$, and therefore the most probable configuration is the one with the vanishing cosmological constant $\Lambda = 0$.

Therefore, unimodular gravity theory provides not only the interpretation of the cosmological constant as an initial condition, but also the reason why the cosmological constant should be zero. Unimodular gravity theory can also provide an alternative solution to the strong CP problem [10].

Considering these developments in unimodular gravity theory, as a possible solution to the cosmological constant problem in gravity physics and also to other fine-tuning problems in particle physics, it seems imperative to consider its supersymmetric generalization, namely to construct unimodular supergravity with local supersymmetry. In this paper, we take the first step toward this direction, namely we present a unimodular supergravity theory with the unit determinant of the metric tensor. We present the lagrangian formulation of unimodular supergravity, in which the metric has a unit determinant as a field equation aided by lagrange multiplier fields. We confirm the closure of the gauge algebra, based on the universal notation in superspace [11], in any arbitrary space-time dimensions $D$ $(1 \leq D \leq 11)$ in which ordinary supergravity theory exists [12]. We use the superspace Bianchi identities [11] in order to show the closure of the gauge algebra holds universally, as long as the ordinary supergravity theory is possible.

Some readers may develop a question about the real necessity of unimodular formulations for supergravity theories: Since the cosmological constant vanishes so long as supersymmetry is not broken, why do we need to impose the unimodular condition to avoid the usual fine-tuning of the cosmological constant? To answer this question, we mention examples of non-vanishing cosmological constants in certain supergravity theories, such as massive ten-dimensional $(D = 10)$ type IIA supergravity [13] which has a free parameter proportional to the cosmological constant, even though the action is invariant under proper supersymmetry transformations. In this paper, we consider the fact that certain supergravity theories have non-zero cosmological constant as a free parameter, but is still invariant under its proper supersymmetry transformations. In other words, we take the standpoint that a unimodular formulation for the cosmological constant is still important even in a supergravity theory.

## 2. Dimension-Independent Formulation of Unimodular Gravity

We start with the review of lagrangian formulation of non-supersymmetric unimodular gravity in arbitrary space-time dimensions $D$. We then try to formulate unimodular supergravity in an arbitrary $D$, as long as ordinary superspace formulation [11] is possible [12].

As the basic principle, we start with the condition that the determinant of the metric tensor is a constant. However, just for simplicity, we choose this constant to be unity. Then the question is how one can impose such a condition on the ordinary Einstein's general relativity, hopefully from a constraint lagrangian. Suppose the total lagrangian $\mathcal{L}$ is composed of three terms $\mathcal{L}_R$, $\mathcal{L}_M$ and $\mathcal{L}_\Lambda$, where $\mathcal{L}_R$ is the standard Hilbert lagrangian, $\mathcal{L}_M$ is a general matter lagrangian whose details are not crucial here, and $\mathcal{L}_\Lambda$ is the constraint lagrangian introduced in order to fix the determinant of the metric tensor:[3]

$$\mathcal{L} \equiv \mathcal{L}_R + \mathcal{L}_M + \mathcal{L}_\Lambda = +\tfrac{1}{4} e^{-1} R(e) + \mathcal{L}_M + \Lambda(e^{-1} - 1) \ . \tag{2.1}$$

---

[3]We are using the signature $(\eta_{mn}) = \text{diag.}(+, -, -, \cdots, -)$. The inverse power for the determinant $e$ in $\mathcal{L}_C$ is due to the definition of $e \equiv \det(e_a{}^m)$ complying with the notation in [11]. Accordingly, the scalar curvature $R(e) \equiv R(e, \phi(e))$ complies with the definition of the Lorentz connection $\phi_{ma}{}^b$ in [11], and the Hilbert lagrangian has the opposite sign to the most common notation [14].

The $\Lambda \equiv \Lambda(x)$ is a real scalar auxiliary field, and $e \equiv \det(e_a{}^m)$. Here we are using the notation in [11] generalized to $D$-dimensional space-time [12][15], namely the indices $m, n, \cdots = 0, 1, \cdots, D-1$ are for curved coordinates, while $a, b, \cdots = 0, 1, \cdots, D-1$ are for local Lorentz coordinates. In general, the matter lagrangian $\mathcal{L}_M$ contains fermionic fields with the vielbeins $e_a{}^m$. The $R(e)$ is the scalar curvature in terms of the Levi-Civita connection $\{{}^\rho_{\mu\nu}\}$ with no torsion, and all the possible torsion terms are separated in $\mathcal{L}_M$. The gravitational field equation is

$$e\frac{\delta \mathcal{L}}{\delta e_m{}^a} = -\tfrac{1}{2}[\,R_a{}^m(e) - \tfrac{1}{2}e_a{}^m R(e) - \kappa T_a{}^m - 2e_a{}^m \Lambda\,] \doteq 0 ~, \quad (2.2)$$

where $\kappa$ is the gravitational coupling, and the symbol $\doteq$ represents a field equation, distinguished from algebraic identities. The $T_a{}^m$ is the usual energy-momentum tensor $\kappa T_a{}^m \equiv +2e\delta\mathcal{L}_M/\delta e_m{}^a$. Note that the scalar field $\Lambda \equiv \Lambda(x)$ enters in the gravitational field equation (2.2), as if it *were* the cosmological 'constant'. Now taking the trace of (2.2) yields

$$\Lambda \doteq \tfrac{1}{2D}\Big[\tfrac{2-D}{2}R(e) - \kappa T\Big] ~, \quad (2.3)$$

where, as usual, $T \equiv T_m{}^m$. Using (2.3) back in (2.2), we get

$$R_{mn}(e) - \tfrac{1}{D}g_{mn}R(e) \doteq \kappa(T_{mn} - \tfrac{1}{D}g_{mn}T) ~. \quad (2.4)$$

Needless to say, this expression covers the familiar case of $D=4$ in [2]. Now the usual technique is to take the covariant divergence of (2.4) to get

$$\begin{aligned}
&\nabla_n[\,R_m{}^n(e) - \tfrac{1}{D}\delta_m{}^n R(e) - \kappa T_m{}^n + \kappa\tfrac{1}{D}\delta_m{}^n T\,] \\
&= \tfrac{1}{2}\nabla_m R(e) - \tfrac{1}{D}\nabla_m R(e) - \kappa\nabla_n T_m{}^n + \kappa\tfrac{1}{D}\nabla_m T \\
&\doteq -\tfrac{1}{D}\nabla_m\Big[\tfrac{2-D}{2}R(e) - \kappa T\Big] \\
&\doteq -2\nabla_m \Lambda \doteq 0 ~.
\end{aligned} \quad (2.5)$$

Here $\nabla_m$ has the Christoffel connection $\{{}^r_{mn}\}$ only in terms of the vielbein with no torsion. As usual, use is made of the Bianchi identity and the matter field equation

$$\nabla_n R_m{}^n(e) \equiv +\tfrac{1}{2}\nabla_m R(e) ~, \quad (2.6a)$$

$$\nabla_n T_m{}^n \doteq 0 ~, \quad (2.6b)$$

together with the form for $\Lambda$ in (2.3). Eq. (2.6b) is the energy-momentum conservation, which is valid as long as $\mathcal{L}_M$ is invariant under general coordinate transformations. Eq. (2.5) implies nothing other than the constancy of $\Lambda$, and therefore the $\Lambda$-term in (2.2) can be regarded as the cosmological constant in the gravitational field equation. In other words, in this formulation, the value of the cosmological constant is understood as an 'initial condition' instead of 'fine-tuning' performed by hand [2][7][6]. Finally, the $\Lambda$-field equation $\delta\mathcal{L}/\delta\Lambda \doteq 0$ yields the unimodular condition

$$e \doteq 1 \quad . \tag{2.7}$$

Thus we have a lagrangian formulation in which the unimodular condition is automatically derived from the total lagrangian.

### 3. Dimension-Independent Unimodular Supergravity

We now generalize this to supergravity in dimensions $D$ $(1 \leq D \leq 11)$ [12], as long as it allows an invariant lagrangian in superspace formulation [11].

Let $\mathcal{L}$ be the total lagrangian composed of the usual supergravity and matter multiplets part $\mathcal{L}_0$ which is locally invariant up to a total divergence, and the constraint lagrangian $\mathcal{L}_\mathrm{C}$:

$$\mathcal{L} \equiv \mathcal{L}_0 + \mathcal{L}_\mathrm{C} \quad , \tag{3.1a}$$

$$\mathcal{L}_\mathrm{C} \equiv \mathcal{L}_\Lambda + \mathcal{L}_\rho \equiv \Lambda(e^{-1} - 1) + e^{-1}\rho^{\underline{\alpha}}[i(\gamma^a \psi_a)_{\underline{\alpha}} - T_{\underline{\alpha}b}{}^b] \quad . \tag{3.1b}$$

The spinorial auxiliary field $\rho^{\underline{\alpha}}$ is needed as a 'superpartner' of $\Lambda$. Here the *underlined* spinorial indices $\underline{\alpha}, \underline{\beta}, \cdots$ include all the possible internal indices, such as those for $Sp(1)$, $SO(N)$ or dottedness for chiralities, *etc.* [12][15]. In our notation, the multiplication by the $\gamma$-matrices satisfies $(\gamma^a \psi_a)_{\underline{\beta}} = (\gamma^a)_{\underline{\beta}}{}^{\underline{\gamma}} \psi_{a\underline{\gamma}} = -(\gamma^a)_{\underline{\beta\gamma}} \psi_a{}^{\underline{\gamma}}$.[4] The $T_{\underline{\alpha}b}{}^c$ in (3.1) is a supertorsion component in superspace appearing in the general supersymmetry transformation of the vielbein

$$\delta_Q e_a{}^m = +i(\overline{\epsilon}\gamma^m \psi_a) - \epsilon^{\underline{\gamma}} T_{\underline{\gamma}a}{}^b e_b{}^m \quad , \tag{3.2}$$

as is derived easily from eq. (5.6.28) in [11], with the universal constraint $T_{\underline{\alpha\beta}}{}^c = i(\gamma^c)_{\underline{\alpha\beta}}$ in superspace. (Cf. (4.1) below). For $T_{\alpha b}{}^b$ in (3.1) or (3.2), we take the $\theta = 0$ sector as is usually expressed by the symbol | [11], but we omit this symbol consistently throughout in this paper. In most formulations of supergravity in diverse dimensions [16], the second term in (3.2) is usually absent, but we keep this term just for completeness. The $T_{\alpha b}{}^c$ can be generally composed of fundamental superfields, but its detailed structure is not crucial here.

---

[4]There may well be some subtlety about the multiplication of the $\gamma$-matrices, depending on $D$, in which the spinorial metric is not the antisymmetric charge-conjugation matrix $C_{\underline{\alpha\beta}}$ but just the Kronecker's delta $\delta_{\underline{\alpha\beta}}$ [15]. Even though we do not go into the details of such a subtlety in this paper, our results will be general for $\forall D$.

Needless to say, $\mathcal{L}_0$ contains the Hilbert lagrangian $+(1/4)R(e)$, so that the general structure (2.1) is still valid even for supergravity lagrangian (3.1), when all the fields (including $\rho^{\underline{\alpha}}$) other than the vielbein are separated from $\mathcal{L}_0 + \mathcal{L}_\rho$ and included into the 'matter' lagrangian $\mathcal{L}_M$. This is also the reason why we have put the factor $e^{-1}$ in $\mathcal{L}_\rho$ for its regular behavior as a spinor under the general coordinate transformations.

The supersymmetry transformation rules for the new fields $\Lambda$ and $\rho^{\underline{\alpha}}$ are

$$\delta_Q \Lambda = -e\, \delta_Q e^{-1} = \epsilon^{\underline{\alpha}}[\, i(\gamma^a \psi_a)_{\underline{\alpha}} - T_{\underline{\alpha}b}{}^b\,]\Lambda \ , \tag{3.3a}$$

$$\delta_Q \rho^{\underline{\alpha}} = e\epsilon^{\underline{\alpha}}\Lambda - e\rho^{\underline{\alpha}} \delta_Q e^{-1} = e\epsilon^{\underline{\alpha}}\Lambda + \rho^{\underline{\alpha}} \epsilon^{\underline{\beta}}[\, i(\gamma^a \psi_a)_{\underline{\beta}} - T_{\underline{\beta}c}{}^c\,] \ , \tag{3.3b}$$

while other component fields in $\mathcal{L}_0$ transform in the usual way. Some remarks are to be made here. First, we note that our supersymmetry transformation rule must be also constrained, such that the condition $e \doteq 1$ is satisfied. Second, using (3.3), we get the supersymmetry transformation of the condition $e^{-1} \doteq 1$ (2.7) as

$$e \delta_Q e^{-1} = -\epsilon^{\underline{\alpha}}[\, i(\gamma^a \psi_a)_{\underline{\alpha}} - T_{\underline{\alpha}c}{}^c\,] \doteq 0 \implies i(\gamma^a \psi_a)_{\underline{\alpha}} - T_{\underline{\alpha}b}{}^b \doteq 0 \ . \tag{3.4}$$

Third, the supersymmetry transformation of (3.4) itself should also vanish:

$$\delta_Q[\, i(\gamma^a \psi_a)_{\underline{\alpha}} - T_{\underline{\alpha}b}{}^b\,] \doteq 0 \ . \tag{3.5}$$

The explicit form of this can be seen in (4.2). In this sense, our supersymmetry transformation rule is 'semi-on-shell', namely all the conditions related to the unimodular condition $e \doteq 1$ to be respected by the 'constrained' supersymmetry transformation. The word 'semi' is used, because we do not use field equations other than those related to the unimodular condition (2.7).

We now confirm the invariance of the total lagrangian $\mathcal{L}_0$ under supersymmetry $\delta_Q$:

$$\begin{aligned} \delta_Q \left(\mathcal{L}_0 + \mathcal{L}_C\right) &= \delta_Q \mathcal{L}_C \\ &\doteq -e(\delta_Q e^{-1})\Lambda(e^{-1}-1) + \Lambda \delta_Q e^{-1} \\ &\quad + (\delta_Q e^{-1})\rho^{\underline{\alpha}}[\, i(\gamma^a \psi_a)_{\underline{\alpha}} - T_{\underline{\alpha}b}{}^b\,] + e^{-1}[\, e\epsilon^{\underline{\alpha}}\Lambda - e\rho^{\underline{\alpha}}(\delta_Q e^{-1})\,][\, i(\gamma^a \psi_a)_{\underline{\alpha}} - T_{\underline{\alpha}b}{}^b\,] \\ &= +e(\delta_Q e^{-1})\Lambda + \Lambda[\, i(\gamma^a \psi_a)_{\underline{\alpha}} - T_{\underline{\alpha}b}{}^b\,] = 0 \ , \end{aligned} \tag{3.6}$$

where we have used (3.5) and the fact that $\delta_Q \mathcal{L}_0 = 0$ up to a total divergence is taken for granted. Our result is universal and applicable to any supergravity theory that allows a superspace formulation [11] and lagrangian formulation [16].

As for the field equations in our unimodular supergravity, these are exactly parallel to (2.2) - (2.7) for the non-supersymmetric case. This is because, as was also mentioned, the structure of the total lagrangian (3.1) with local supersymmetry is exactly the same as that in (2.1), when all the gravitino-dependent terms and other fields (including $\rho^{\underline{\alpha}}$) in $\mathcal{L}_0$ and $\mathcal{L}_\rho$ are separated and collected into $\mathcal{L}_{\rm M}$. In particular, this $\mathcal{L}_{\rm M}$ is to have the regular invariance under general coordinate transformations, in order to have the energy-momentum conservation (2.6b). For example, the factor $e^{-1}$ in $\mathcal{L}_\rho$ becomes important, because if this factor were absent, then $\rho^{\underline{\alpha}}$ would have to transform as a spinor 'density' instead of a regular spinor, which would modify the energy-momentum conservation (2.6b). Eventually, the field equations (2.2) - (2.7) are still valid, yielding the same condition of $\Lambda \doteq$ const. with $e \doteq 1$. Hence we emphasize that our total lagrangian (3.1) is valid as the locally supersymmetric unimodular supergravity lagrangian, and is applicable to any supergravity theory with lagrangian formulation [12][16] in arbitrary dimensions $D$ ($1 \leq D \leq 11$).

## 4. Closure of Gauge Algebra in Unimodular Supergravity

This section contains the most non-trivial part of our formulation. We confirm the closure of gauge algebra on our supersymmetric unimodular conditions (2.7), (3.4) and (3.5). Note that since the result in this section is algebraic, it is more general than the lagrangian formulation of the previous section, as long as a given supergravity theory allows superspace formulation [11].

The supersymmetry transformation of (2.7) yielded an additional condition (3.4). We first review the derivation of (3.4) based on the general formulae (5.6.28) in [11]: First, we get

$$\delta_Q e_a{}^m = -\epsilon^{\underline{\beta}} T_{\underline{\beta} a}{}^b e_b{}^m - \epsilon^{\underline{\beta}} \psi_a{}^{\underline{\gamma}} T_{\underline{\gamma}\underline{\beta}}{}^b e_b{}^m = -\epsilon^{\underline{\gamma}} [ + i(\gamma^m)_{\underline{\gamma}\underline{\delta}} \psi_a{}^{\underline{\delta}} + T_{\underline{\gamma} a}{}^b e_b{}^m ] \ . \qquad (4.1)$$

Next, multiplying this by $e^{-1} e_m{}^a$, we get $\delta_Q e^{-1}$ in (3.4). The supersymmetry transformation of (3.5) in turn yields the additional condition

$$i(\gamma^a D_a(\widehat{\phi})\epsilon)_{\underline{\beta}} + \epsilon^{\underline{\gamma}} \nabla_{\underline{\gamma}} T_{\underline{\beta} c}{}^c$$
$$+ i(\gamma^a)_{\underline{\beta}\underline{\gamma}} \Big[ \epsilon^{\underline{\delta}} T_{\underline{\delta} a}{}^{\underline{\gamma}} + \epsilon^{\underline{\delta}} T_{\underline{\delta} a}{}^b \psi_b{}^{\underline{\gamma}} + \epsilon^{\underline{\delta}} \psi_a{}^{\underline{\epsilon}} T_{\underline{\epsilon}\underline{\delta}}{}^{\underline{\gamma}} + \epsilon^{\underline{\delta}} \psi_a{}^{\underline{\epsilon}} T_{\underline{\epsilon}\underline{\delta}}{}^b \psi_b{}^{\underline{\gamma}} \Big] \doteq 0 \ . \qquad (4.2)$$

The Lorentz connection $\widehat{\phi}$ contains what is called the $\psi$-torsion [17] as well as other torsion components via $T_{ab}{}^c$: $\widehat{\phi}_{ma}{}^b \equiv \widehat{\phi}_{ma}{}^b(e, \psi, T)$. These will be given explicitly in (4.9). Eq. (4.2) is easily confirmed by the expressions for $\delta_Q \psi_a{}^{\underline{\gamma}}$ given in [11] and $\delta_Q T_{\underline{\beta} c}{}^c = -\epsilon^{\underline{\gamma}} \nabla_{\underline{\gamma}} T_{\underline{\beta} c}{}^c$, as

$$\begin{aligned}
0 \doteq\ & \delta_Q [i(\gamma^a \psi_a)_{\underline{\beta}} - T_{\underline{\beta} c}{}^c] = -i(\gamma^a)_{\underline{\beta}\underline{\gamma}} (\delta_Q \psi_a{}^{\underline{\gamma}}) - \delta_Q T_{\underline{\beta} c}{}^c \\
=\ & -i(\gamma^a)_{\underline{\beta}\underline{\gamma}} \Big[ D_a(\widehat{\phi}) \epsilon^{\underline{\gamma}} - \epsilon^{\underline{\delta}} T_{\underline{\delta} a}{}^{\underline{\gamma}} - \epsilon^{\underline{\delta}} T_{\underline{\delta} a}{}^b \psi_b{}^{\underline{\gamma}} - \epsilon^{\underline{\delta}} \psi_a{}^{\underline{\epsilon}} T_{\underline{\epsilon}\underline{\delta}}{}^{\underline{\epsilon}} T_{\underline{\epsilon}\underline{\delta}}{}^{\underline{\gamma}} - \epsilon^{\underline{\delta}} \psi_a{}^{\underline{\epsilon}} T_{\underline{\epsilon}\underline{\delta}}{}^b \psi_b{}^{\underline{\gamma}} \Big] \\
& + \epsilon^{\underline{\gamma}} \nabla_{\underline{\gamma}} T_{\underline{\beta} c}{}^c \ . \qquad (4.3)
\end{aligned}$$

The condition (4.2) dictates the space-time dependence of the parameter $\epsilon^{\underline{\alpha}}$.

On the other hand, under translations, the determinant $e^{-1}$ transforms as

$$\delta_P e^{-1} = -\xi^m \partial_m e^{-1} - e^{-1} \partial_m \xi^m = -\partial_m(e^{-1}\xi^m) \ , \tag{4.4}$$

where $\xi^m$ is the parameter for the translation. The non-trivial confirmation now is to see whether the parameter $\xi^m$ satisfies the condition

$$\partial_m(e^{-1}\xi^m) \doteq 0 \ , \tag{4.5}$$

to be consistent with our first condition (2.7). In particular, the parameter $\xi^m$ is to be identified with that arising from the commutator of two supersymmetries $[\delta_Q(\epsilon_1), \delta_Q(\epsilon_2)] = \delta_P(\xi^m)$ [11]:

$$\xi^a \equiv i(\overline{\epsilon}_1 \gamma^a \epsilon_2) \ . \tag{4.6}$$

Eq. (4.5) is further rewritten via (4.6) as

$$i[\, D_m(\widehat{\phi})(e^{-1}e_a{}^m)\,](\overline{\epsilon}_1 \gamma^a \epsilon_2) + [\,ie^{-1}(\overline{\epsilon}_1 \gamma^a D_a(\widehat{\phi})\epsilon_2) - {\scriptstyle(1 \leftrightarrow 2)}\,] \doteq 0 \ . \tag{4.7}$$

The first term here can be further rewritten in terms of supercovariant anholonomy coefficients [11]:

$$C_{ab}{}^c \equiv (e_a{}^n \partial_n e_b{}^m - e_b{}^n \partial_n e_a{}^m)e_m{}^c + i(\overline{\psi}_a \gamma^c \psi_b) - \psi_{[a|}{}^{\underline{\alpha}} T_{\underline{\alpha}|b]}{}^c + T_{ab}{}^c$$
$$\equiv C_{ab}{}^c(e) + C_{ab}{}^c(\psi, T) \ , \tag{4.8}$$

related to $\widehat{\phi}_{abc} \equiv \widehat{\phi}_{abc}(e, \psi)$ as

$$\widehat{\phi}_{abc} = +\tfrac{1}{2}(C_{abc} - C_{acb} + C_{cba}) \ . \tag{4.9}$$

The derivative factor in the first term in (4.7) can be re-expressed as

$$eD_m(\widehat{\phi})(e^{-1}e_a{}^m) = -C_{ab}{}^b(\psi, T)$$
$$= -i(\overline{\psi}_a \gamma^b \psi_b) + \psi_a{}^{\underline{\alpha}} T_{\underline{\alpha} b}{}^b - \psi_b{}^{\underline{\alpha}} T_{\underline{\alpha} a}{}^b - T_{ab}{}^b$$
$$= -\overline{\psi}_a{}^{\underline{\beta}}[i(\gamma^b \psi_b)_{\underline{\beta}} - T_{\underline{\beta} b}{}^b] - \psi_b{}^{\underline{\alpha}} T_{\underline{\alpha} a}{}^b - T_{ab}{}^b$$
$$\doteq -\psi_b{}^{\underline{\alpha}} T_{\underline{\alpha} a}{}^b - T_{ab}{}^b \ . \tag{4.10}$$

Here the first term in the penultimate line has vanished due to our condition (4.4). Now using (4.10) in (4.7), we can confirm (4.5) via (4.7) as

$$0 \stackrel{?}{=} (\psi_b{}^{\underline{\alpha}} T_{\underline{\alpha} b}{}^b \xi^a - T_{ab}{}^b \xi^a) + [\,i(\overline{\epsilon}_1 \gamma^a D_a(\widehat{\phi})\epsilon_2) - {\scriptstyle(1 \leftrightarrow 2)}\,]$$
$$\doteq -i\psi_b{}^{\underline{\alpha}} T_{\underline{\alpha} a}{}^b (\overline{\epsilon}_1 \gamma^a \epsilon_2) - iT_{ab}{}^b(\overline{\epsilon}_1 \gamma^a \epsilon_2)$$
$$+ \epsilon_1^{\underline{\beta}} \epsilon_2^{\underline{\delta}}\Big[-i(\gamma^a)_{(\underline{\beta}|\underline{\gamma}} T_{|\underline{\delta})a}{}^{\underline{\gamma}} - i(\gamma^a)_{(\underline{\beta}|\underline{\gamma}} T_{|\underline{\delta})a}{}^b \psi_b{}^{\underline{\gamma}}$$
$$- i(\gamma^a)_{(\underline{\beta}|\underline{\gamma}} \psi_a{}^{\underline{\epsilon}} T_{\underline{\epsilon}|\underline{\delta})}{}^{\underline{\gamma}} - i(\gamma^a)_{(\underline{\beta}|\underline{\gamma}} T_{\underline{\epsilon}|\underline{\delta})}{}^b \psi_b{}^{\underline{\gamma}} - \nabla_{(\underline{\beta}} T_{\underline{\delta})c}{}^c\Big] \ . \tag{4.11}$$

Here use is also made of the condition (4.2) to get rid of the derivative term of $\epsilon$. The symbol $\stackrel{?}{=}$ is used, because the equality under question has yet to be confirmed.

In order to simplify the term $\nabla_{\underline{\beta}} T_{\underline{\delta}c}{}^c$ in (4.11), we next use the supertorsion Bianchi-identity

$$\nabla_{(\underline{\beta}} T_{\underline{\delta})c}{}^c + \nabla_c T_{\underline{\beta}\underline{\delta}}{}^c - T_{\underline{\beta}\underline{\delta}}{}^d T_{dc}{}^c - T_{\underline{\beta}\underline{\delta}}{}^{\underline{\epsilon}} T_{\underline{\epsilon}c}{}^c \\ - T_{c(\underline{\beta}|}{}^d T_{d|\underline{\delta})}{}^c - T_{c(\underline{\beta}|}{}^{\underline{\epsilon}} T_{\underline{\epsilon}|\underline{\delta})}{}^c - R_{\underline{\beta}\underline{\delta}c}{}^c \equiv 0 \ . \tag{4.12}$$

Due to the universal constraint $T_{\underline{\alpha}\underline{\beta}}{}^c = i(\gamma^c)_{\underline{\alpha}\underline{\beta}}$ [11], the second term in the first line vanishes. Because of (anti)symmetry of indices, the first term in the second line and the last term also vanish. Eventually we get

$$\nabla_{(\underline{\beta}} T_{\underline{\delta})c}{}^c \equiv T_{\underline{\beta}\underline{\delta}}{}^{\underline{\epsilon}} T_{\underline{\epsilon}c}{}^c + i(\gamma^c)_{(\underline{\beta}|\underline{\epsilon}} T_{c|\underline{\delta})}{}^{\underline{\epsilon}} + i(\gamma^d)_{\underline{\beta}\underline{\delta}} T_{dc}{}^c \ , \tag{4.13}$$

which, after the substitution into the last term in (4.11), simplifies the latter as

$$0 \stackrel{?}{=} -i\psi_b{}^{\underline{\alpha}} T_{\underline{\alpha}a}{}^b (\bar{\epsilon}_1 \gamma^a \epsilon_2) - i T_{ab}{}^b (\bar{\epsilon}_1 \gamma^a \epsilon_2)$$
$$+ \epsilon_1^{\underline{\beta}} \epsilon_2^{\underline{\delta}} [ -i(\gamma^a)_{(\underline{\beta}|\underline{\gamma}} T_{|\underline{\delta})a}{}^{\underline{\gamma}} - i(\gamma^a)_{(\underline{\beta}|\underline{\gamma}} T_{|\underline{\delta})a}{}^b \psi_b{}^{\underline{\gamma}}$$
$$- i(\gamma^a)_{(\underline{\beta}|\underline{\gamma}} \psi_a{}^{\underline{\epsilon}} T_{\underline{\epsilon}|\underline{\delta})}{}^{\underline{\gamma}} + (\gamma^a)_{(\underline{\beta}|\underline{\gamma}} \psi_a{}^{\underline{\epsilon}} (\gamma^b)_{\underline{\epsilon}|\underline{\delta})} \psi_b{}^{\underline{\gamma}}$$
$$- T_{\underline{\beta}\underline{\delta}}{}^{\underline{\epsilon}} T_{\underline{\epsilon}c}{}^c - i T_{c(\underline{\beta}|}{}^{\underline{\epsilon}} (\gamma^c)_{\underline{\epsilon}|\underline{\delta})} - i(\gamma^b)_{\underline{\beta}\underline{\delta}} T_{dc}{}^c ] \tag{4.14a}$$
$$\stackrel{.}{=} \epsilon_1^{\underline{\beta}} \epsilon_2^{\underline{\delta}} [ + i(\gamma^a)_{\underline{\beta}\underline{\delta}} \psi_b{}^{\underline{\gamma}} T_{\underline{\gamma}a}{}^b - i(\gamma^a)_{(\underline{\beta}|\underline{\gamma}} T_{|\underline{\delta})a}{}^b \psi_b{}^{\underline{\gamma}}$$
$$- i(\gamma^a)_{(\underline{\beta}|\underline{\gamma}} \psi_a{}^{\underline{\epsilon}} T_{\underline{\epsilon}|\underline{\delta})}{}^{\underline{\gamma}} - i T_{\underline{\beta}\underline{\delta}}{}^{\underline{\epsilon}} (\gamma^c \psi_c)_{\underline{\epsilon}} ] \tag{4.14b}$$
$$\stackrel{.}{=} \tfrac{1}{2} \epsilon_1^{\underline{\alpha}} \epsilon_2^{\underline{\beta}} \psi_b{}^{\underline{\gamma}} [ i(\gamma^a)_{(\underline{\alpha}\underline{\beta}} T_{\underline{\gamma})a}{}^b - i(\gamma^b)_{(\underline{\alpha}|\underline{\delta}} T_{|\underline{\beta}\underline{\gamma})}{}^{\underline{\delta}} ] \ , \tag{4.14c}$$

after various cancellation among like terms. In (4.14a), the last term in the first line and the last term in the last line cancel each other, so do the first term in the second line and the second terms in the last line. The last term in the third line vanishes due to the antisymmetry $(\gamma^a \psi_b)_{(\underline{\beta}|} (\gamma^b \psi_a)_{|\underline{\delta})} \equiv 0$. From (4.14b) to (4.14c), use is also made of the condition (4.4) replacing $T_{\underline{\epsilon}c}{}^c$ by $i(\gamma^c \psi_c)_{\underline{\epsilon}}$. Now our last task is to show that (4.14c) vanishes. Fortunately, this can be easily done by the use of another supertorsion Bianchi-identity

$$\tfrac{1}{2} \nabla_{(\underline{\alpha}} T_{\underline{\beta}\underline{\gamma})}{}^b - \tfrac{1}{2} T_{(\underline{\alpha}\underline{\beta}|}{}^d T_{d|\underline{\gamma})}{}^b - \tfrac{1}{2} T_{(\underline{\alpha}\underline{\beta}|}{}^{\underline{\delta}} T_{\underline{\delta}|\underline{\gamma})}{}^b - \tfrac{1}{2} R_{(\underline{\alpha}\underline{\beta}\underline{\gamma})}{}^b \equiv 0 \ , \tag{4.15}$$

where the first and last terms vanish. Eq. (4.15) implies that (4.14c) vanishes identically, and therefore (4.5) vanishes as desired, *via* (4.11) and (4.7). This concludes our proof of (4.5) for the parameter (4.2).

Before concluding this section, we briefly consider the closure of supersymmetry on our new fields $\Lambda$ and $\rho^{\underline{\alpha}}$. First, notice that the *on-shell* closure on these fields is easier to handle than the lagrangian invariance, because here we can use field equations. Second, as has been mentioned, once the *semi-on-shell* condition (3.4) is considered, both of the fields $\Lambda$ and $\rho^{\underline{\alpha}}$ do not transform. In other words, the commutator $[\delta_Q(\epsilon_1), \delta_Q(\epsilon_2)]$ vanishes, when acting on both of these fields. At first sight, this sounds puzzling, because this also means the absence of translation generated on both of these fields. However, note that the field $\Lambda$ is to be a constant after all, so that its translation is required to vanish. As for the field $\rho^{\underline{\alpha}}$, it can be completely gauged away by an appropriate local supersymmetry in (3.3b), when $\Lambda \doteq$ const. In other words, any transformation of $\rho^{\underline{\delta}}$, including the usual translation, can be re-absorbed into a new supersymmetry parameter. Therefore the vanishing of the commutator $[\delta_Q(\epsilon_1), \delta_Q(\epsilon_2)]$ on both of these new fields poses no problem for closure of supersymmetry.

## 5. $N=1$ Unimodular Supergravity in $D=4$ as An Example

Once we have established our general formulation of unimodular supergravity, it is easier to look into some explicit examples. Here we give an example of old minimal supergravity [18] in $D=4$.

Complying with the superspace notation so far, we give the superspace constraints for supertorsions and supercurvatures of $D=4$, $N=1$ old minimal supergravity [18] for the component field content $(e_a{}^m, \psi_a{}^{\underline{\alpha}}, S, P, A_m)$:

$$T_{\underline{\alpha}\underline{\beta}}{}^c = +i(\gamma^c)_{\underline{\alpha}\underline{\beta}} \ , \tag{5.1a}$$

$$T_{\underline{a}b}{}^{\underline{\gamma}} = -\tfrac{i}{6}(\gamma_b)_{\underline{\alpha}}{}^{\underline{\gamma}}S - \tfrac{1}{6}(\gamma_5\gamma_b)_{\underline{\alpha}}{}^{\underline{\gamma}}P - \tfrac{i}{3}(\gamma_5\gamma_b\gamma^c)_{\underline{\alpha}}{}^{\underline{\gamma}}A_c \ , \tag{5.1b}$$

$$\nabla_{\underline{\alpha}}S = +\tfrac{1}{2}(\gamma^{ab})_{\underline{\alpha}\underline{\gamma}}T_{ab}{}^{\underline{\gamma}} = +\tfrac{i}{2}(\gamma^m\mathcal{R}_m)_{\underline{\alpha}} \ , \tag{5.1c}$$

$$\nabla_{\underline{\alpha}}P = -\tfrac{i}{2}(\gamma_5\gamma^{ab})_{\underline{\alpha}\underline{\gamma}}T_{ab}{}^{\underline{\gamma}} = +\tfrac{1}{2}(\gamma_5\gamma^m\mathcal{R}_m)_{\underline{\alpha}} \ , \tag{5.1d}$$

$$\nabla_{\underline{\alpha}}A_b = +\tfrac{3}{4}(\gamma_5\gamma_b{}^{cd})_{\underline{\alpha}}{}^{\underline{\gamma}}T_{cd}{}^{\underline{\gamma}} - \tfrac{1}{2}(\gamma_5\gamma_b\gamma^{cd})_{\underline{\alpha}\underline{\gamma}}T_{cd}{}^{\underline{\gamma}} = +\tfrac{3i}{2}(\gamma_5\mathcal{R}_b)_{\underline{\alpha}} - \tfrac{i}{2}(\gamma_5\gamma_b\gamma^c\mathcal{R}_c)_{\underline{\alpha}} \ , \tag{5.1e}$$

$$T_{ab}{}^c = +\tfrac{2}{3}\epsilon_{ab}{}^{cd}A_d \ , \tag{5.1f}$$

with $\mathcal{R}_{a\underline{\alpha}} \equiv -(i/2)(\gamma_a{}^{bc})_{\underline{\alpha}\underline{\beta}}T_{bc}{}^{\underline{\beta}}$, while all other remaining supertorsion components $T_{\underline{\alpha}b}{}^c$, $T_{\underline{\alpha}\underline{\beta}}{}^{\underline{\gamma}}$ between the dimensionality $0 \leq d \leq 1$ are zero. Here the *underlined* spinorial indices are for the four-component spinors: $\underline{\alpha}$ = 1, 2, 3, 4. The component invariant lagrangian $\mathcal{L}_0$ of supergravity corresponding to (3.1a) [18] and $\mathcal{L}_C$ of (3.1b) are

$$\begin{aligned}\mathcal{L}_0 + \mathcal{L}_C = \ & +\tfrac{1}{4}e^{-1}R(e,\phi(e)) - \tfrac{1}{2}\epsilon^{mnrs}\overline{\psi}_m\gamma_5\gamma_n D_r(\phi(e))\psi_s - \tfrac{1}{6}e^{-1}(S^2+P^2-A_m^2) \\ & + \Lambda(e^{-1}-1) + ie^{-1}\rho^{\underline{\alpha}}(\gamma^m\psi_m)_{\underline{\alpha}} \ , \end{aligned} \tag{5.2}$$

due to the absence of $T_{\underline{\alpha}b}{}^c$. The supersymmetry transformation rules corresponding equations to (3.2), (3.3a) and (3.3b) are simple, because we simply drop the last terms with $T_{\underline{\alpha}b}{}^c$.

By adding also the transformations of $\psi_a{}^{\underline\alpha}$, $S$, $P$ and $A_a$, we complete the supersymmetry transformation rule as

$$\delta_Q e_a{}^m = +i(\overline\epsilon\gamma^m\psi_a) \ , \tag{5.3a}$$

$$\delta_Q \psi_m{}^{\underline\alpha} = D_m(\phi(e,\psi))\epsilon^{\underline\alpha} - \tfrac{i}{6}(\gamma_m\epsilon)^{\underline\alpha}S + \tfrac{1}{6}(\gamma_5\gamma_m\epsilon)^{\underline\alpha}P + \tfrac{i}{3}(\gamma_5\epsilon)^{\underline\alpha}A_m - \tfrac{i}{6}(\gamma_5\gamma_m{}^n\epsilon)^{\underline\alpha}A_n \ , \tag{5.3b}$$

$$\delta_Q S = -\tfrac{i}{2}(\overline\epsilon\gamma^m\mathcal{R}_m) \ , \qquad \delta_Q P = -\tfrac{1}{2}(\overline\epsilon\gamma_5\gamma^m\mathcal{R}_m) \ ,$$

$$\delta_Q A_m = -\tfrac{3i}{2}(\overline\epsilon\gamma_5\mathcal{R}_m) + \tfrac{i}{2}(\overline\epsilon\gamma_5\gamma_m\gamma^n\mathcal{R}_n) \ , \tag{5.3c}$$

$$\delta_Q \Lambda = +i(\overline\epsilon\gamma^m\psi_m) \ , \tag{5.3d}$$

$$\delta_Q \rho^{\underline\alpha} = +e\epsilon^{\underline\alpha}\Lambda + ie\rho^{\underline\alpha}(\overline\epsilon\gamma^m\psi_m) \ . \tag{5.3e}$$

The previous invariance confirmation (3.6) for our total lagrangian $\mathcal{L} \equiv \mathcal{L}_0 + \mathcal{L}_{\mathrm{C}}$ is performed in exactly the same way here, and also the closure on all the fields as in section 4 as well.

Note that even though we are using here the 'off-shell' formulation of $D=4$, $N=1$ supergravity with the old minimal multiplet [18], the closure of gauge algebra related to the unimodular condition, *i.e.*, those equations in section 5, are 'semi on-shell'. This seems inevitable, as long as we impose the unimodular condition from outside, even if it is implied by 'auxiliary' multiplier fields $\Lambda$ and $\rho^{\underline\alpha}$ at the lagrangian level.

## 6. $N=1$ Unimodular Supergravity in $D=11$ as Another Example

As another instructive and useful application, we look at $N=1$ supergravity in $D=11$. There is a slight difference in this system compared with the previous $D=4$ case.

In our unimodular supergravity formulation, as some careful readers may have noticed, we have seen that all the supertorsions/supercurvatures constraints in superspace have not been modified, but there are additional constraints on fields such as (2.7) and (3.4), or constraints on the supersymmetry parameter (4.2). In component language, this is equivalent to the fact that all the transformation rules for the original fields, such as $e_a{}^m$ and $\psi_a{}^{\underline\alpha}$ are not modified formally, but these fields are more constrained than before by the constraints (2.7) and (3.4), *etc.* The only new transformation rule is for the new fields $\Lambda$ and $\rho^{\underline\alpha}$. Therefore considering superspace Bianchi identities [11][19], there will be no 'modifications' for the original field equations for the original fields.[5] The only new ingredient is the constraints (2.7) or (3.4) on the original fields, together with the constraint (4.2) on the supersymmetry parameter.

Considering these points, it is now clear that in the case of $N=1$ supergravity in $D=11$, there will be no cosmological constant possible. This is because all the original field equations including also the gravitational one are maintained, allowing no cosmological constant. To be more specific, the gravitational field equation implied by the original Bianchi identities [11][19]

$$R_{mn} = -\tfrac{1}{3}(F_{mrst}F_n{}^{rst} - \tfrac{1}{12}g_{mn}F_{rstu}F^{rstu}) \ , \tag{6.1}$$

stays the exactly the same even in the unimodular case with no cosmological constant. The unimodular condition $e \doteq 1$ of (2.7) follows from the $\Lambda$-field equation $\delta \mathcal{L}/\delta \Lambda \doteq 0$ consistently with supersymmetry, while (6.1) forces not only $\Lambda \doteq$ const. but also $\Lambda \doteq$ const. $\doteq 0$. Namely, we get $\Lambda$ to be zero exactly, maintaining the original field equations consistent under supersymmetry.

---
[5]We use here the words 'constraints' distinguished from 'modifications', because all the original form of field equations are formally maintained.

From this viewpoint, our formulation of unimodular supergravity is more meaningful, when the value of cosmological constant is not determined by local supersymmetry itself, such as in $1 \leq D \leq 10$.

## 7. Effects of Superpartner Fields to Vacuum Functional

The vacuum functional (1.1) in unimodular gravity [9] had been derived, ignoring any contributions to $Z$ by matter fields other than graviton. However, in our unimodular supergravity, there are superpartner matter fields, such as gravitino or other bosonic as well as fermionic fields. Therefore, it is reasonable to ask about their possible contributions to the vacuum functional. To answer this question, we point out that the derivation of (1.1) is based on the assumption that the effects of 'backgrounds' (but *not* fluctuations) of matter fields are negligible [20]. As a matter of fact, the vacuum functional (1.1) is derived from

$$\begin{aligned} Z &= \int d\mu(\Lambda) \ \exp\left[-S_\Lambda(\overline{g}_{\mu\nu}, \overline{\phi})\right] \\ &= \int d\mu(\Lambda) \ \exp\left[-S_\Lambda(\overline{g}_{\mu\nu}, 0)\right] \\ &= \int d\mu(\Lambda) \ \exp\left(\tfrac{3\pi}{G\Lambda}\right) \ , \end{aligned} \tag{7.1}$$

where $S_\Lambda(\overline{g}_{\mu\nu}, 0) = -3\pi/(\Lambda G)$ is the action for a four-sphere background metric $\overline{g}_{\mu\nu}$. In other words, it is the 'background' values $\overline{\phi}$ (distinguished from their original values $\phi$) of all the matter fields that might contribute to $Z$. In the non-supersymmetric case [9], it is assumed that $\overline{\phi}$ are all vanishing, and therefore, the second line in (7.1) follows. In our present paper, we rely on the same assumption for superpartner fields in our unimodular supergravity in a given $D$-dimensional space-time, such as gravitino, fermionic matter or higher-rank bosonic fields. Since the purpose of this paper is to consider the vanishing of a cosmological constant within the given $D$-dimensional space-time, instead of compactifications with Freund-Rubin type background values for bosonic fields, our assumption here seems quite legitimate. To conclude, the presence of superpartner fields in unimodular supergravity theory does not upset the good feature with vacuum functional inherent in unimodular non-supersymmetric gravity.

## 8. Concluding Remarks

In this paper, we have shown that the unimodular supergravity theory can be formulated in any space-time dimensions $D\ (1 \leq D \leq 11)$, in which ordinary supergravity theory exists [12], based on the universal notation in superspace [11]. We have presented the supergravity lagrangian that generates the unimodular determinant condition as a field equation of a multiplier field. We have confirmed the invariance of our lagrangian under local supersymmetry up to total divergence. We have seen that the non-trivial closure of the gauge algebra is confirmed with the help of Bianchi identities in superspace, in a highly sophisticated but universal way applicable to any supergravity theory [11][16], independent of the space-time dimensions $D\ (1 \leq D \leq 11)$.

In section 2, we have presented a lagrangian formulation, assuming that the basic supergravity theory allows a lagrangian in $D\ (1 \leq D \leq 11)$. Therefore those supergravity theories allowing no lagrangian formulations, such as type IIA supergravity in 10D are excluded in section 2. However, armed with the algebraic closure confirmed in section 4, we can also include those supergravity theories even without lagrangian formulations. This is another advantage of our analysis of gauge algebra in section 4.

As for the possible contributions by superpartner matter fields to the vacuum functional (1.1), we have understood that the usual assumption for the background values (but not their fluctuations) of matter fields to be vanishing, and will not affect the vacuum functional (7.1). This is reasonable for fermionic fields such as gravitino, and other bosonic fields as well, within a given $D$-dimensional space-time, in which a conventional supergravity theory can be formulated. Therefore, the good feature of unimodular non-supersymmetric gravity has been inherited to our unimodular supergravity.

Some readers may be wondering about other quantum behaviour of unimodular supergravity from a general viewpoint. For example, one might think that the condition of unit determinant of the metric would introduce unphysical degrees of freedom, as it is equivalent to a nontrivial gauge condition, and therefore unitarity of the theory is doubtful. Furthermore, at the loop level new divergences might appear as densities of arbitrary weight. Such a worry, however, is not appropriate. This is because we have established supergravity formulation of unimodular gravity, which is compatible with superstring whose quantum behaviour is much better, or supposed to be finite to all orders, compared with general relativity. In this context, we re-emphasize the importance of investigating supergravity formulation of unimodular gravity. It is not just pure curiosity that we need to establish supergravity formulation for unimodular gravity, but it is only after supergravity formulation is established, that we have a good control of the quantum behaviour of unimodular gravity theory.

The success of the universal formulation of unimodular supergravity indicates that the concept of unit determinant for metric tensor has fundamental significance, compatible with supersymmetry. In other words, the necessity of such a formulation in order to understand the vanishing cosmological constant as an initial condition instead of fine-tuning by hand, is compatible also with supersymmetry, which is another important concept in particle physics.

We believe that our results presented in this paper will provide a good working ground for future study of unimodular supergravity/supersymmetry models that may provide solutions for other 'fine-tuning' problems both in gravity and particle physics.

We are grateful to J. Gates, Jr. for helpful discussions. Special acknowledgment is due to M. Luty for initial collaboration on the subject, and for suggesting to include the explicit examples of $D = 4$ and $D = 11$ unimodular supergravity theories. Final acknowledgement is due to the referee of this paper who helped us for clarifying important aspects of unimodular supergravity.